\documentclass[aps,prd,floatfix,superscriptaddress,preprintnumbers,nofootinbib,preprint,tightenlines,longbibliography]{revtex4-1}
\pdfoutput=1

\usepackage{amsmath, amssymb, amsfonts, graphicx, mathrsfs, verbatim, float, slashed, bbm, color, xcolor, xspace, enumerate, url, bbding, multirow, outlines,upgreek}
\usepackage[english]{babel}
\usepackage[colorlinks=true,linkcolor=black,citecolor=galacticcenterbubblegum]{hyperref}
\definecolor{galacticcenterbubblegum}{rgb}{0.8,0, 0.8}

\newcommand{\beq}{\begin{equation}}
\newcommand{\bea}{\begin{eqnarray}}
\newcommand{\eeq}{\end{equation}}
\newcommand{\eea}{\end{eqnarray}}
\newcommand{\bal}{\begin{align}}
\newcommand{\eal}{\end{align}}

\newcommand{\Lc}{\mathcal{L}}

\newcommand{\Mc}{\mathcal{M}}

\newcommand{\Z}{\mathbb{Z}}

\newcommand{\Aa}{{\bf A1}}
\newcommand{\Ab}{{\bf A2}}
\newcommand{\Ba}{{\bf B1}}
\newcommand{\Bb}{{\bf B2}}
\newcommand{\C}{{\bf C}}

\newcommand{\MeV}{~\mathrm{MeV}}
\newcommand{\GeV}{~\mathrm{GeV}}
\newcommand{\TeV}{~\mathrm{TeV}}

\newcommand{\Tmax}{~T_\text{max}}
                    
  \definecolor{darklightsabergreen}{rgb}{0.0, .49, 0.06}

\begin{document}

\title{Neutron Decay to a Non-Abelian Dark Sector}

\author{Fatemeh Elahi}
\affiliation{School of Particles and Accelerators, Institute for Research in Fundamental Sciences IPM, Tehran, Iran}
\author{Mojtaba Mohammadi Najafabadi}
\affiliation{School of Particles and Accelerators, Institute for Research in Fundamental Sciences IPM, Tehran, Iran}
\vspace{2.4cm}


 \begin{abstract}

According to the Standard Model (SM), we expect to find a proton for each decaying neutron. However, the experiments counting the number of decayed neutrons and produced protons have a disagreement. This discrepancy suggests that neutrons might have an exotic decay to a Dark Sector (DS). In this paper, we explore a scenario where neutrons decay to a dark Dirac fermion $\chi$ and a non-abelian dark gauge boson $W'$. We discuss the cosmological implications of this scenario assuming DS particles are produced via freeze-in. In our proposed scenario, DS has three portals with the SM sector: (1) the fermion portal coming from the mixing of the neutron with $\chi$, (2) a scalar portal, and (3) a non-renormalizable kinetic mixing between photon and dark gauge bosons which induces a vector portal between the two sectors. We show that neither the fermion portal nor the scalar portal should contribute to the production of the particles in the early universe. Specifically, we argue that the maximum temperature of the universe must be low enough to prevent the production of $\chi$ in the early universe. In this paper, we rely on the vector portal to connect the two sectors, and we discuss the phenomenological bounds on the model. The main constraints come from ensuring the right relic abundance of dark matter and evading the neutron star bounds. When dark gauge boson is very light, measurements of the Big Bang Nucleosynthesis impose a stringent constraint as well. 
 \end{abstract}

\maketitle

 \section{Introduction}
 \label{sec:intro}
Even though the Standard Model (SM) of particle physics can explain almost all observed phenomena, we are certain there exits physics beyond the SM. One of the most prominent questions in the particle astrophysics community is the nature and origin of dark matter (DM). So far, we have not observed any unambiguous detection of DM. However, numerous experimental anomalies may be a hint of DM interaction with the SM. One of these experiments is the measurements of the neutron lifetime. 

Due to the importance of neutrons as one of the main building blocks of luminous matter and one of the key role players in the formation of light elements in the early universe, there have been several experiments that attempt to find the lifetime of the neutrons~\cite{Mampe:1993an,Robson:1951fq,Serebrov:2004zf,Pichlmaier:2010zz,Steyerl:2012zz,Czarnecki:2018okw,1309.2623,Wietfeldt:2011suo,Byrne:1990ig,Yue2013}. In the SM, we expect the branching ratio of a neutron to a proton, an electron, and a neutrino ($n \to p + e+\bar \nu_e$) to be $100\%$. In an experiment known as the bottle experiment~\cite{Mampe:1993an,Robson:1951fq,Serebrov:2004zf,Pichlmaier:2010zz,Steyerl:2012zz,Czarnecki:2018okw,1309.2623,Wietfeldt:2011suo}, ultracold neutrons are stored for a time comparable to the neutron lifetime, then the remaining neutrons are counted. This experiment finds the total decay width or equivalently the lifetime of the neutrons. Their finding is $ \tau_n^{\text{bottle}} = 879.6 \pm 0.6 s$. In another experiment known as the beam experiment~\cite{Byrne:1990ig,Yue2013}, the number of produced protons are counted, and their finding has been announced to be $\tau_{n\to p + ...}^{\text{beam}} = 888.0 \pm 2.0 s$. The lifetime of neutrons in these two experiments differ by $7.6 s$ by $4\sigma$, which constitutes about $1\%$ branching ratio of the neutron. The aforementioned discrepancy may be the result of an exotic decay of neutrons to the dark sector. Due to the close mass of neutrons and protons and their intimate structures, the easiest way to ensure an exotic decay of a neutron and the stability of protons is to assume the total mass of the exotic decay of neutron $M_f$ is greater than the mass of proton and electron: $ m_n > M_f > m_p - m_e$. Furthermore, any baryon number violating process is severely constrained~\cite{Miura:2016krn,Phillips:2014fgb,Goldman:2019dbq,Grossman:2018rdg,Leontaris:2018blt,Berezhiani:2017jkn,Gardner:2017szu,Aaij:2017inn,Aharmim:2017jna,Fomin:2017aiz,Hewes:2017xtr,Liu:2016oik,Frost:2016qzt,1708.01259,1512.05359}. Therefore, we are led to consider scenarios where neutrons can decay to a new degree of freedom that has a baryonic charge. 

Numerous studies have explored different possibilities~\cite{1801.01124,Davoudiasl:2014gfa,Cline:2018ami,Barducci:2018rlx,Ivanov:2018uuk,Fornal:2018ubn,Bringmann:2018sbs,Fornal:2018mhk,Grinstein:2018ptl,Berezhiani:2018udo,Berezhiani:2018eds,Fabbrichesi:2019ema,Fornal:2019olr,Garani:2018kkd,Keung:2020teb,Wietfeldt:2020ock,Dubbers:2018kgh,Wietfeldt:2018upi}. The most minimalistic scenario discussed in the literature is  $n \to \chi \gamma$, where $\chi$ is a fermionic DM that has a baryonic charge. If we assume $m_p < m_\chi < m_n$, we expect $E_\gamma < 1.572 \MeV$~\cite{1801.01124}. However, experimental measurements disfavor this scenario~ \cite{Tang:2018eln,Klopf:2019afh} for a photon with an energy in the range $0.782 \MeV< E_\gamma < 1.664 \MeV$ up to $2.2$ sigma. Softer photons remain unexplored. Extending the scenario to include another $\phi$ with mass $m_\phi < 1.572 \MeV$ is also another possibility.  An important ingredient for both of these scenarios is a mixing between the neutron and $\chi$, which has an effective Lagrangian of 
\beq
\Lc_{\text{eff}} = \bar \chi (i \slashed{D}- m_\chi) \chi - \bar n (i \slashed \partial - m_n + \mu_n \sigma^{\mu\nu} F_{\mu\nu}) n - \delta m \bar n_R \chi_L.
\eeq

To resolve the neutron lifetime discrepancy, $\delta m$ is expected to be about $10^{-11} \MeV$~\cite{1801.01124}. A conversion of neutrons to $\chi$ with such strength has important consequences in the equation of states of Neutron Stars (NSs) ~\cite{Cline:2018ami}.  That is the decay of neutrons in the NS cause the equation of state (pressure and energy density of the neutron star) to alter significantly, and thereby affect the mass and volume of NSs. Specifically, if neutron and $\chi$ are in chemical equilibrium, due to the less interaction of $\chi$ comparatively, the conversion of neutrons to $\chi$ leads to a lower pressure. Integrating the Tolman-Oppenheimer-Volkoff equations~\cite{1802.08244,1802.08244,1802.08427,Tolman:1939jz, Oppenheimer:1939ne,1010.5788,1101.1921,Fabbrichesi:2019ema,Garani:2018kkd,Keung:2020teb}, one can find the maximum mass of an NS as a function of its radius, and the upper limit is in contradiction with the properties of some of the neutron stars observed~\cite{Cline:2018ami}. The simplest solution is to consider DM scenarios that have repulsive self-interaction and a repulsive interaction with neutrons. That is to have a vector mediator, e.g. a dark photon. 

In Ref.~\cite{Cline:2018ami}, the authors considered the decay of $ n \to \chi A'$. To ensure the theory is consistent with the observation of dense NSs with radius $2 M_\odot$, we need $m_{A'}/g_D \lesssim (45-60) \MeV$, where $g_D$ is the gauge coupling of the $U(1)_D$.  In this setup, we also have a mixing between the dark gauge boson and the SM photon: 
\beq
\Lc = - \frac{\epsilon}{4} F'_{\mu \nu} F^{\mu \nu},
\eeq
which induces dark photon - electromagnetic current  interaction with a coupling proportional to $\epsilon$. The authors of Ref.~\cite{Cline:2018ami} did an extensive phenomenological study of this scenario, and showed that the parameter space for $ m_{A'} < 2m_e$ is severely constrained, while the case where $m_{A'} > 2 m_e$ is slightly better. One of the main constraints comes from the era of Big Bang Nucleosynthesis (BBN), which requires dark photon to decay early enough that it won't inject much energy during BBN. One way to escape this constraint is producing the dark sector (DS) particles via freeze-in. In this paper, we explore this possibility and show that a $\chi$ that is a stable DM candidate and can justify the neutron decay experiments will necessarily over-close the universe. Thereby, we must require the maximum temperature of the universe to be low enough that $\chi$ is not produced in the early universe. To explain the relic abundance of DM, we need to rely on the dark photon or the scalar, both of which are unstable particles.  Hence, in this paper, instead of a dark $U(1)_D$, we consider a dark non-abelian gauge  $SU(2)_D$, because even though it does not have any more free parameters, the extra degrees of freedom helps with explaining the two observations of DM and neutron decay. Furthermore, for the freeze-in scenario to work, we should employ very small kinetic mixing, and this is more justified in the non-abelian kinetic mixing because of its non-renormalizable nature. The main differences between our work and Ref.~\cite{Cline:2018ami} are the followings:
\begin{itemize}
\item[--] In this paper, DS has a gauge $SU(2)_D$ rather than a gauge $U(1)_D$.
\item[--] We assume the relic abundance of DS particles is through freeze-in. 
\item [--] We turn off the scalar portal between the two sectors. More specifically, in the potential term  $\lambda_{\phi H} |\phi|^2|H|^2$ -- with $\phi$ being the scalar responsible for the spontaneous breaking of the $SU(2)_D$  and $H$ being the SM Higgs -- we take $\lambda_{\phi H} = 0$. We argue that the radiative correction is very suppressed, and thus our assumption is justifiable. This choice of $\lambda_{\phi H}$ has important consequences for the relic abundance of DM.  
\end{itemize}

 The organization of the paper is as follows: In section~\ref{sec:model} we explain the model and introduce the degrees of freedom as well as the free parameters in the theory. Section~\ref{sec:pheno} is denoted to the phenomenology of the model including the constraints from NSs, the neutron decay experiments, and the cosmological constraints which is discussed in Section~\ref{sec:cosmo}. Direct Detection and Collider Constraints are explored in Section~\ref{sec:DD}, and finally, the concluding remarks are presented in the Conclusion~\ref{sec:conclusion}.

 \section{Model}
 \label{sec:model}

Let us assume Dark Sector has a gauge $SU(2)_D$ that is spontaneously broken by a doublet $\phi$:
\beq
\phi = \begin{pmatrix} \frac{1}{\sqrt{2}}(G^1_\phi + i G_\phi^2)\\  \varphi + v_\phi+ i G_\phi^3\end{pmatrix},
\eeq
where $G^i_\phi$ are the Goldstone bosons, which become the longitudinal component of the gauge bosons, and $v_\phi$ is vacuum expectation value (\textit{vev}) of $\phi$. To ensure $\phi$ indeed acquires \textit{vev}, we require its potential to have the following form: 
\beq
V(\phi,H) = - \mu_\phi^2 |\phi|^2 + \lambda_\phi |\phi|^4 - \mu_H^2 |H|^2 + \lambda_H |H|^4 + \lambda_{\phi H} |\phi|^2 |H|^2, 
\eeq 
with $\mu_\phi^2 > 0$. Since neutron is a fermion, the decay of a neutron to DS particles compels us to include fermionic degrees of freedom. Thereby, we introduce a Dirac fermion $\chi$ transforming as a doublet under $SU(2)_D$: $\chi^T = (\chi_1, \chi_2)$. Since there are severe constraints on the baryon number violating models~\cite{Miura:2016krn,Phillips:2014fgb}, we assume $\chi$ has a baryon charge of $+1$. The effective Lagrangian becomes
\begin{align}
\Lc_{NP} =& - \frac{1}{4} W^{'a}_{\mu \nu}W^{'a,\mu \nu} + i \bar \chi  (\slashed{D} + m_\chi) \chi + |D_\mu \phi|^2 + \eta \bar \chi_L \phi n  +h.c.\nonumber\\
& + C_Y \phi^\dagger \tau^a \phi W^{'a}_{\mu \nu}F^{\mu \nu} + \tilde C_{Y1} \phi^\dagger \tau^a \phi W^{'a}_{\mu \nu}\tilde F^{\mu \nu}+ \tilde C_{Y2} \phi^\dagger \tau^a \phi \tilde W^{'a}_{\mu \nu} F^{\mu \nu}+h.c \nonumber\\
&- V(\phi,H),
\label{eq:Lag}
\end{align}
where $ W^{'a}_{\mu \nu}$ is the field strength tensor of $SU(2)_D$ and $ \tilde  W^{'a}_{\mu \nu}= \epsilon_{\mu \nu \alpha \beta}  W^{'a, \alpha \beta} $ is the dual of the field tensor, and $ D_\mu = \partial_\mu - i g_D \tau^a W^a_\mu$. Once $\phi$ acquires \textit{vev}, a mixing between the neutron and $\chi_1$ is induced, which results in the conversion of neutrons to $\chi$ and other dark sector particles. 

The Lagrangian terms written in the second line of Eq.~\ref{eq:Lag} are the non-abelian kinetic mixing between $SU(2)_D$ gauge bosons' and the photon's field tensor. Due to the presence of $\phi$ in these terms the kinetic mixing with the CP-odd component is not a total derivative, and thus contribute to the action. For simplicity, we assume $\tilde C_Y \equiv \tilde C_{Y1} = \tilde C_{Y2}$. Note that the kinetic mixing terms are dimension 6 operators and therefore have an inverse mass-squared dimension. The suppressed mass dimension means there is a small couplings of the dark gauge bosons with SM particles.  

Note that with this setup, another effective Lagrangian term $\bar \chi_L \phi \pi p$ can be written as well. This term may lead to proton decay, $\chi_1$ decay, or neither depending on the mass spectrum. We must assume $m_p < m_\chi < m_n$ to ensure the stability of proton as the lightest fermion charged under the baryon symmetry $U(1)_B$. If $m_\chi > m_p + m_\pi$, $\chi_1$ may decay. However, in this paper, we fix $m_\chi$ such that $\chi_1$ is also a stable particle. \footnote{Even though after $\phi$ acquires a vev, there is a slight mass splitting between $\chi_1$ and $\chi_2$, this mass splitting is negligible compared with $m_{\chi}$. Therefore, for the rest of the paper, we will assume both $\chi_1$ and $\chi_2$ have mass $m_{\chi}$ and we will use $\chi$ to refer to both of them.} 

After $SU(2)_D$ Spontaneous Symmetry Breaking (SSB), $W'$ and $\phi$ get a mass proportional to $v_\phi$:
$$ m_{_{W'}} = g_D v_\phi \hspace{0.7 in} m_\phi  = \sqrt{\lambda_\phi} v_\phi.$$
At low energies there is a residual $\Z_2$ symmetry remaining from the broken $SU(2)_D$. Under the $\Z_2$ symmetry, $W^{'\pm}$ and $\chi_2$ are odd, and the rest of the particles are even. It has been shown that to explain the neutron decay anomaly ($n \to \chi W^{'\pm}$) and yet be safe from the NS constraint, we necessarily need to have $m_\chi \gg m_{_{W'}}$. Therefore, $W^{'\pm}$ are the lightest particles charged under $\Z_2$, and thus are stable. 
This is while $W'_3$ mixes with photon after $\phi$ gets a \textit{vev}, and thus it can decay (e.g, $ W'_3 \to e^+ e^-$ if $m_{_{W'}} > 2m_e$ and $ W'_3 \to 3 \gamma$  for lighter $W'$.) Similarly, due to the $\phi  W'_3  W'_3$ coupling we can have $\phi \to \gamma \gamma$ decay. It is worth mentioning that $W'_3$ and/or $\phi$ could be long-lived DM candidates.

The free parameters in this model are the following:
\begin{itemize}
\item  \indent masses : $ m_{\chi}, ~ m_\phi,  ~ v_\phi$. 
\item \indent  couplings : $\eta , g_D, \lambda_{\phi H},$ and $C_Y, \tilde C_Y$ with dimensions proportional to $[M^{-2}]$.
\end{itemize}

To find the cosmological constraints on the model, we need to briefly discuss the UV completion of the model. This is very similar to the model suggested in Ref.~\cite{Cline:2018ami}: two color triplet scalars with hypercharge $1/3$ ( $\Phi_1$ and $\Phi_2$) are introduced, where $\Phi_1$ is also a doublet of $SU(2)_D$. Thus, the UV Lagrangian can simply be written as 
\beq
\Lc_{\text{UV}} = \lambda_1\bar d^aP_{L\chi} \Phi_{1a} + \lambda_2 \epsilon^{abc}\bar u^c_a P_R d_b \Phi_{2c} + \mu_1 \Phi_{1a} \Phi_2^{*a}\phi,
\eeq
where in the effective theory 
\beq
 \eta = \frac{ \beta v_\phi  \lambda_1 \lambda_2}{m_{\Phi_1}^2 m_{\Phi_2}^2},
\eeq
with $\beta$ being the factor derived from confinement of quarks to neutrons, and its value $\beta = 0.014 \GeV^3$ is taken from Lattice QCD simulations~\cite{Aoki:2017puj}. Dijet searches at CMS~\cite{1703.09986} and ATLAS~\cite{1712.02332} push the masses of $\phi_i$ to greater than $1 \TeV$. 

Since $\Phi_i$, with $i = 1,2$ are charged under $SU(3)_c$, we expect their number density in the early universe to match that of photons (e.g., we expect them to be in thermal equilibrium with thermal bath). Through their couplings with the dark sector, the production of $\chi$ and subsequently $\phi$ and $W'$ should occur in abundance. As shown in~\cite{Cline:2018ami}, such set up leads to severe constraints from CMB~\cite{Cline:2013fm,1705.02358,1805.03656}, BBN~\cite{Hufnagel:2017dgo}, and the \textit{Fermi}-LAT observation of gamma rays from dwarf spheroidal galaxies~\cite{Drlica-Wagner:2015xua,0704.0261}. The summary of these constraints in presented here:
\begin{itemize}
\item[--] Once $\chi$ becomes non-relativistic, it can only annihilate to $W' W'$ and $\phi \phi$ efficiently. Therefore, if $\chi$ is in thermal equilibrium in the early universe, the abundant production of $W'$ and $\phi$ becomes inevitable. On the other hand, $W'_3$ and $\phi$ can only decay after $\phi$ acquires a \textit{vev}, which roughly occurs around $60 \MeV$ \footnote{This value is the maximum value allowed from the NS constraint}, and it is extremely close to BBN. The decay of $W'_3$ and $\phi$ near the BBN disturbs the Hubble rate and thus it significantly alters the production of light nuclei by diluting the baryon-photon ratio as well as causing photodissociation of the nuclei. 
\item[--]  Similarly, decays near and during recombination will distort the CMB temperature fluctuations and thus there are severe constraints on a model with light $W'$ from CMB as well. 
\item[--] For $m_{W'} \ll m_\chi$, the annihilation of $\chi\chi$ to $W'W'$ is Sommerfeld at low velocities, which leads to an enhanced annihilation cross-sections in spheroidal galaxies and at the time of recombination. Therefore, it is crucial that we do not have much $\chi$ in the universe. 
\end{itemize} 

Thereby, in this paper, we explore another avenue. We assume dark sector particles start with zero abundance in the early universe and they get produced through freeze-in mechanism~\cite{Hall:2009bx, Elahi:2014fsa,Bernal:2019mhf}. Consequently, in our set-up, we need the maximum temperature $\Tmax$ to be smaller than $m_{\Phi_{_i}}$ so that they are not produced in the early universe. If the color multiplets are not produced, then the production of $\chi$ is greatly reduced. 

\begin{figure}[h!]
\centering 
\includegraphics[width=0.45\textwidth, height=0.19\textheight]{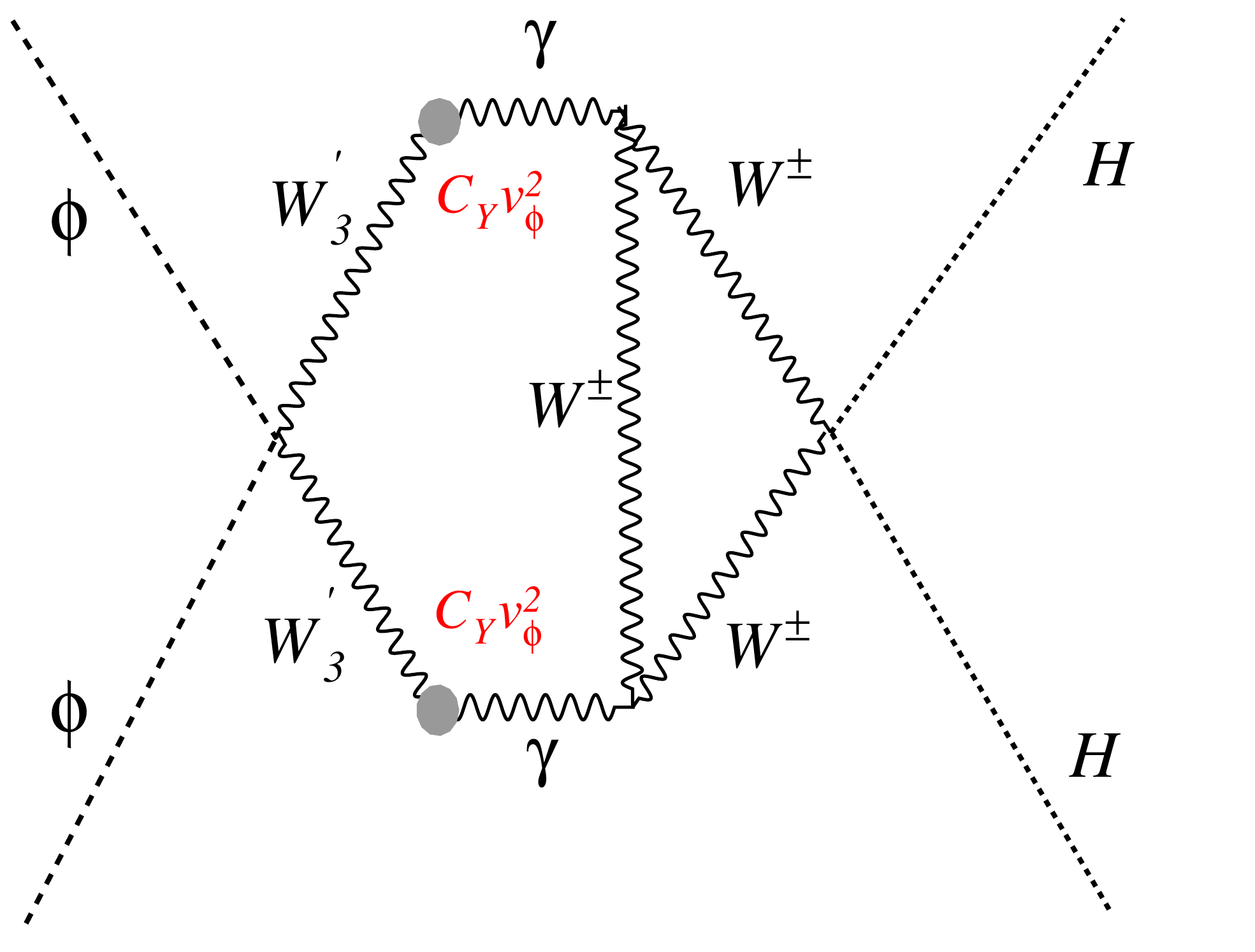}
\caption{ The radiative corrections to $\lambda_{\phi H}$. As illustrated this coupling is suppressed by two loops as well as $C_Y^2 v_\phi^4$, and thus it is very small. Therefore, if we let $\lambda_{\phi H} =0$, we can easily make sure the radiative corrections to $\lambda_{\phi H}$ stay negligible. }
\label{fig:lambda}
\end{figure}

The portals between the dark sector and the SM sector are via (1) the Higgs portal with a strength proportional to $\lambda_{\phi H}$, (2) the kinetic mixing governed by $C_Y v_\phi^2 $, and (3) the effective mixing between $\chi$ and neutron which is $\eta v_\phi$. For a successful freeze-in scenario, we need an extremely weak connection between the dark sector and the SM sector. Since $C_Y$ and $\eta$ are due to non-renormalizable interactions, we can justifiably assign them small values\footnote{The value of $\eta$ is governed by the neutron decay anomaly. Since $\eta$ is derived from non-renormalizable interaction, we expect its value to be small. }. This argument becomes more non-trivial for $\lambda_{\phi H}$, which in general can take any value $\lesssim 1$. If we want to assign $\lambda_{\phi H}$ a small value, we must make sure that this choice is safe from loop corrections. The radiative correction to $\lambda_{\phi H}$ comes from $\phi\phi\gamma\gamma$ vertex, which only opens up after $\phi$ acquires a \textit{vev} and even then it is suppressed -- the coupling is proportional to $g_D C_Y^2 v_\phi^4$.  Fig.\ref{fig:lambda} shows one of the leading diagrams to radiative correction to $\lambda_{\phi H}$, and as it is illustrated, in addition to the $C_Y^2 v_\phi^4$ suppression, it is two loops suppressed. Therefore, if the value of $\lambda_{\phi H}$ is small at tree level, it does not get amplified much at loop levels. For simplicity, in this work, we assume $\lambda_{\phi H} = 0$. 

Having closed the Higgs portal, now we need to discuss the evolution of dark sector particles in the early universe, and how much they contribute to the relic abundance of the total DM.  In the following section, we discuss the phenomenological constraints on each of these parameters including the ideal spot that explains the neutron decay anomaly and yields the correct relic abundance of DM. Even though the number of free degrees of parameters is large, the numerous experimental and observations bounds on these parameters forces us to live in a small region of the parameter space. 
 
 \section{phenomenology}
 \label{sec:pheno} 
 
One of the most important bounds on this model comes from NSs, where the conversion of the neutron to $\chi$ can have significant consequences. If neutrons and $\chi$ are in chemical equilibrium, it is favorable for the neutrons to convert into $\chi$, which due to its almost non-interacting nature, results in a lower pressure in the neutron stars. By integrating the Tolman-Oppenheimer-Volkoff equation, one finds the maximum mass with respect to the NS's radius falls below the largest observed mass. Our scenario falls in the category that there is a repulsive interaction between $\chi$ and the neutrons and thus can be safe from this constraint as long as $m_{_{W'}}/g_D < 60 \MeV$~\cite{Cline:2018ami,1802.08244}.

Another important restriction in this model comes from neutron decay. As it has been shown in Refs.~\cite{Tang:2018eln, Cline:2018ami}, the decay of neutron\footnote{ The decay of $n \to \chi \gamma$ can occur via $\bar n \sigma_{\mu \nu} \phi^\dagger \chi  F^{\mu \nu}$. However, because of the null search for monochromatic photon~\cite{Tang:2018eln} and exacerbating the tension of the axial coupling of the neutron~\cite{1802.01804}, we expect this coupling to be very small and negligible in this study.}  to $\chi W'$ and $\chi \phi$ are 

\begin{align}
\Gamma_{n \to \chi _{W'}}  \simeq \frac{\eta^2 (m_n - m_\chi)^2 }{8 \pi m_n } \left[1-  \left(\frac{m_{W'}}{m_n- m_\chi}\right)^2\right]^{3/2}\\
\Gamma_{n \to \chi \phi} \simeq \frac{\eta^2 (m_n - m_\chi)^2 }{16  \pi m_n} \left[1-  \left(\frac{m_{\phi}}{m_n- m_\chi}\right)^2\right]^{3/2}.
\label{eq:decaywidth}
\end{align}

We have already discussed that the mass of $\chi$ should satisfy $m_p + m_e < m_\chi < m_n$, in order to both satisfy the neutron decay and yet be safe from stringent proton decay bounds. To have a decay that is kinematically allowed, we must have  $m_\chi + \text{Min}\left[m_{_{W'}}, m_\phi\right] < m_n$. Therefore, let us consider the following benchmarks: 
\begin{itemize}
\item We consider two benchmarks where both $W'$ and $\phi$ are light enough that both decays mentioned in Eq.~\ref{eq:decaywidth} are allowed. For one of these benchmarks, we take $m_\phi > 2 m_{W'}$:
$$m_\chi = 937.992\MeV, \hspace{0.3 in} m_\phi = 1.4 \MeV, \hspace{0.3 in} \text{and} ~m_{_{W'}} = 0.5 \MeV. $$ 
Note that in this benchmark, $\phi$ decays to $W'$, and thus it is not a DM candidate. We will denote this benchmark as $\Aa$. To justify the neutron decay discrepancy, we need $\eta \simeq 2.8 \times 10^{-11}$. 

Another benchmark we choose is when $m_\phi \sim m_{W'}$:
$$m_\chi = 937.992\MeV, \hspace{0.3 in} m_\phi = 1.4 \MeV, \hspace{0.3 in} \text{and}~ m_{_{W'}} = 1.4 \MeV, $$
and we present this benchmark by $\Ab$.  The $\eta$ that explain the neutron decay is $\eta \simeq7.6 \times 10^{-11}$. 
It is worth mentioning that $\phi$, in this benchmark, can decay to two photons. However, depending on $C_Y$, $\phi$ can be a long lived DM candidate. 
 \item Another scenario is when $\phi$ is heavy such that the decay $ n \to \chi \phi$ is not kinematically allowed. In this case, we will also take two different benchmarks; one where $m_\phi > 2 m_{_{W'}}$: 
 $$m_\chi = 937.992\MeV,\hspace{0.3 in} m_\phi = 4  \MeV, \hspace{0.3 in} \text{and} ~m_{_{W'}} = 1.4 \MeV, $$ 
 which we use $\Ba$ to refer to this benchmark. Solving for the $\eta$ that yields $\text{Br} (n \to \chi W') \simeq 1\%$ is $\eta \simeq 9.3 \times 10^{-11}$. 
 
 Another benchmark, satisfies $m_{_{W'}} < m_\phi < 2 m_{_{W'}}$: 
  $$m_\chi = 937.992\MeV,\hspace{0.3 in} m_\phi = 2 .5 \MeV, \hspace{0.3 in} \text{and} ~m_{_{W'}} = 1.4 \MeV. $$ 
This benchmark is presented by $\Bb$. Since the only parameter that has changed is $m_\phi$ and $n \to \chi \phi$ is forbidden, the desired $\eta$ is still $\eta \simeq 9.3 \times 10^{-11}$.
\item We also consider another case where $m_{W'}$ is large enough that the decay of $n \to \chi W'$ is not allowed. However, $\phi$ is light enough that allows the dark decay of neutrons:
  $$m_\chi = 937.992\MeV,\hspace{0.3 in} m_\phi = 1.4  \MeV, \hspace{0.3 in} \text{and} ~m_{_{W'}} = 2.5 \MeV. $$
  This benchmark is referred by $\C$. The desired $\eta$ to justify the neutron decay anomaly is $\eta \simeq 1.3 \times 10^{-10}$. 
\end{itemize}

 \subsection{Cosmology}
 \label{sec:cosmo} 
In this section, we will discuss the cosmological constraints, and we will see satisfying the relic abundance and making sure DM candidates do not over-close the universe gives the most stringent bound for most of our benchmarks. Moreover, BBN, which strongly disfavors new degrees of freedom injecting energy around the formation of nuclei in the early universe, puts important constraints on some of the benchmarks. As mentioned earlier, the observation of large neutron stars excludes part of the parameter space as well. The rest of the constraints from various experiments and observations are also discussed in this section.  
 \subsubsection{Relic Abundance}
 \label{sec:relic} 
We are interested in a scenario where dark sector particles start with zero or negligible abundance and then are slowly produced through their feeble interactions with SM particles. First, we will discuss the production of $\chi$ as it will be important to set the maximum temperature of the universe, then we will investigate the evolution of $W'$ and $\phi$ in the early universe. \\

\small{ \underline{$\chi$ Production}}\\

Up until $\phi$ acquires a \textit{vev},\footnote{we will assume that the temperature at which $\phi$ gets a \textit{vev} is the value of \textit{vev} itself $T_{v_\phi} \sim 60 \MeV$.} the production of $\chi$ is due to $q q \to q \phi \chi$, where $q = u,d$. The Boltzmann equation describing the evolution of $\chi$ number is 
\beq 
\dot n_\chi + 3 H n_\chi = \int \ d \Pi_q\  d\Pi_q\ d\Pi_q\ d\Pi_\phi\ d\Pi_\chi\ (2\pi)^4 (p_{\text{i}} - p_{\text{f}})\ |\Mc|^2_{qq \to q \phi \chi}\  f_q\ f_q,
\label{eq:Bol}
\eeq
where $d\Pi_i = \frac{d^3p_i}{(2\pi)^3 2 E_i} $, and $f_q \sim e^{-E_q/T}$ is the distribution function of the quarks in thermal bath, and $s$ is the canonical Mandelstam variable. We can simplify Eq.~\ref{eq:Bol} for any process that has three final state particles ~\cite{Elahi:2014fsa}:
\beq 
\dot n_\chi+ 3 Hn_\chi = \frac{T}{(4\pi)^6} \int d \Omega \int_{0}^{\infty} ds\  s^{3/2} |\Mc|^2_{qq \to q \phi \chi} K_1 \left(\frac{\sqrt{s}}{T}\right),
\label{eq:2to3}
\eeq
where $K_1(x)$ denotes the modified Bessel function of the second kind. Eq.~\ref{eq:2to3} is in the relativistic limit where the masses of the particles involved are negligible to the temperature. The squared Matrix Element (ME) of  $q q \to q \phi \chi$ in the relativistic limit is
\beq
|\Mc|^2_{q q \to q \phi \chi} \simeq 36  \left(\frac{\eta}{\beta}\right)^2 s^2. 
\eeq
The integral over $s$ in Eq.~\ref{eq:2to3} has a closed form 
\beq
\int_0^{\infty} ds\  s^{(2n+1)/2} K_1\left(\frac{\sqrt{s}}{T}\right)= 4^{n+1} T^{2n+3} \Gamma (n+1) \Gamma (n+2),
\label{eq:intBessel} 
\eeq
for $n > -1$.  Therefore, we can easily calculate the right hand side of Eq.~\ref{eq:Bol}. The left hand side can be converted to yield ($Y \equiv n/S$) with $S$ being the entropy density:
\begin{align} 
Y_\chi &= \int_0^{\Tmax} d T \left\{- \frac{1}{SHT}\left[ \frac{T}{(4\pi)^6} \int d \Omega \int_{0}^{\infty} ds\  s^{3/2} |\Mc|^2_{qq \to q \phi \chi} K_1 \left(\frac{\sqrt{s}}{T}\right)\right]\right\} \nonumber\\
&\simeq \frac{3^5 * 5^2}{ 1.66 \times  \pi^{9} \sqrt{g_\star^{\rho}} g_\star^S} M_{Pl} \left( \frac{\eta  }{\beta}\right)^2 T_{\text{max}}^5 \times \theta(\Tmax - m_\chi) \nonumber\\
& \simeq  \left\{\begin{array}{ccc} 6 \times 10^{-4}  &&\Aa\\ 4.3 \times 10^{-3} && \Ab\\ 6.5 \times 10^{-3} && \Ba\& \Bb\\ 0.013 && \C\end{array} \right. \times \left(\frac{ T_{\text{max}}}{\GeV}\right)^5 \times  \theta(\Tmax - m_\chi)
\end{align}
where the second equality is obtained by using the definitions $S= \frac{2\pi^2 g_\star^ST^3}{45} $ and $H= \frac{1.66 \sqrt{g_\star^\rho} T^2}{M_{Pl}}$, and $\theta(x)$ is the step function that ensures the universe has enough energy to produce $\chi$.  
There is a constraint on the $Y_\chi$ so that it does not over-close the universe:
\beq
Y_{\chi} \leq \frac{ \Omega_{\text{total DM}} \rho_c}{m_\chi  s_0} \simeq 6.6 \times 10^{-11}  \times\left( \frac{ \GeV}{m_\chi}\right),
\eeq 
where for all of our benchmarks, the mass of $\chi$ is fixed to $ m_\chi = 937.992 \MeV$. It is clear that if $\chi$ is produced, it quickly over-closes the universe. Thereby, we require  $ \Tmax < m_\chi$ to prevent the production of $\chi$ in the early universe. In other words, even though $\chi$ is a stable particle, it does not contribute to the relic abundance of the DM in the universe. Notice that these calculations only depend on the coupling of $\chi$ with quarks and thus the results are the same if we had assume a $U(1)_D$ instead of the $SU(2)_D$. \\

\small{\underline{$W'$ and $\phi$ Production}}\\
  
\begin{figure}[h!]
\centering 
\includegraphics[width=0.37\textwidth, height=0.15\textheight]{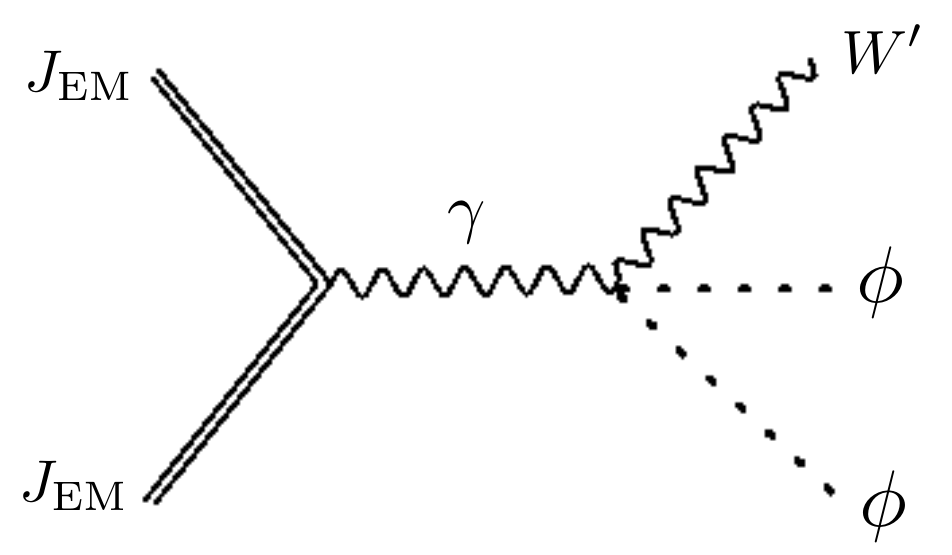}
\caption{The Feynman diagram that leads to the production of $\phi$ and $W'$, where $J_{\text{EM}}$ means the electromagnetic current.}
\label{fig:phiw}
\end{figure}

For the case where $\lambda_{\phi H} = 0$, the main mechanism for the production of $\phi$ and $W'_\mu$ is via the kinetic mixing term. The leading diagram of $W'/\phi$ production for $T > v_\phi$ is shown in Fig.~\ref{fig:phiw}, where $J_{\text{EM}}$ represents any particle (lepton or hadron) that has electromagnetic charge and has a significant abundance at $ T < m_\chi$. The Boltzmann equation governing the number density of $W'$ and $\phi$ is similar to Eq.~\ref{eq:2to3}, and the squared ME for the process of our interest $J_{_{\text{EM}}} J_{_{\text{EM}}} \to \phi  \phi W'$ is the following: 
\beq
|\Mc|^2_{_{ J_{_{\text{EM}}} J_{_{\text{EM}}} \to \phi  \phi W'}}= \frac{ Q^4 C_Y^2 (4 s^2 -5 s (t+u) -5 tu )} {4 s},
\label{eq:MEhighT} 
\eeq
 with $Q$ being charge of the initial state particles, and $s,t,$ and $u$ are the Mandelstam variables. In the limit where $s \gg t, u$, we get $|\Mc|^2_{_{ J_{_{\text{EM}}} J_{_{\text{EM}}} \to \phi  \phi W'}} \to Q^4 C_Y^2 s $.  In Eq.~\ref{eq:MEhighT}, we have let $m_{_{W'}} = m_\phi = 0$, because $\phi$ still has not acquired a \textit{vev}. Note that $W'$ and $\phi$ do not get thermal corrections as well, since they live in a much colder sector. In the SM sector, however, for $ T \gg m{_{\text{SM}}} (T= 0)$, the thermal correction to the mass of particles becomes important. For simplicity, we assume that $m_{_{\text{SM}}} (T) \sim T$. The only exception is for proton, where for $ T < m_p$, we assume $m_p (T) = m_p(T=0)$. Furthermore, we make the reasonable assumption that unstable particles decay more efficiently to SM particles than annihilate to $W'$ and $\phi$.  The yield of $W'$ and $\phi$ coming from Fig.~\ref{fig:phiw} is   
 \beq
 Y_{_{UV_i}}\simeq  g_i \alpha M_{Pl} \frac{3^3 5^2 \times C_Y^2}{ 1.66 \times 2^4  \pi^{8} \sqrt{g_\star^{\rho}} g_\star^S} T_{\text{max}}^3,
 \label{eq:yieldUV} 
 \eeq
 where $i = W'/\phi$ and $g_i$ represents the number of $\phi$ and $W'$ produced.  

\begin{figure}[h!]
\centering 
\includegraphics[width=0.75\textwidth, height=0.3\textheight]{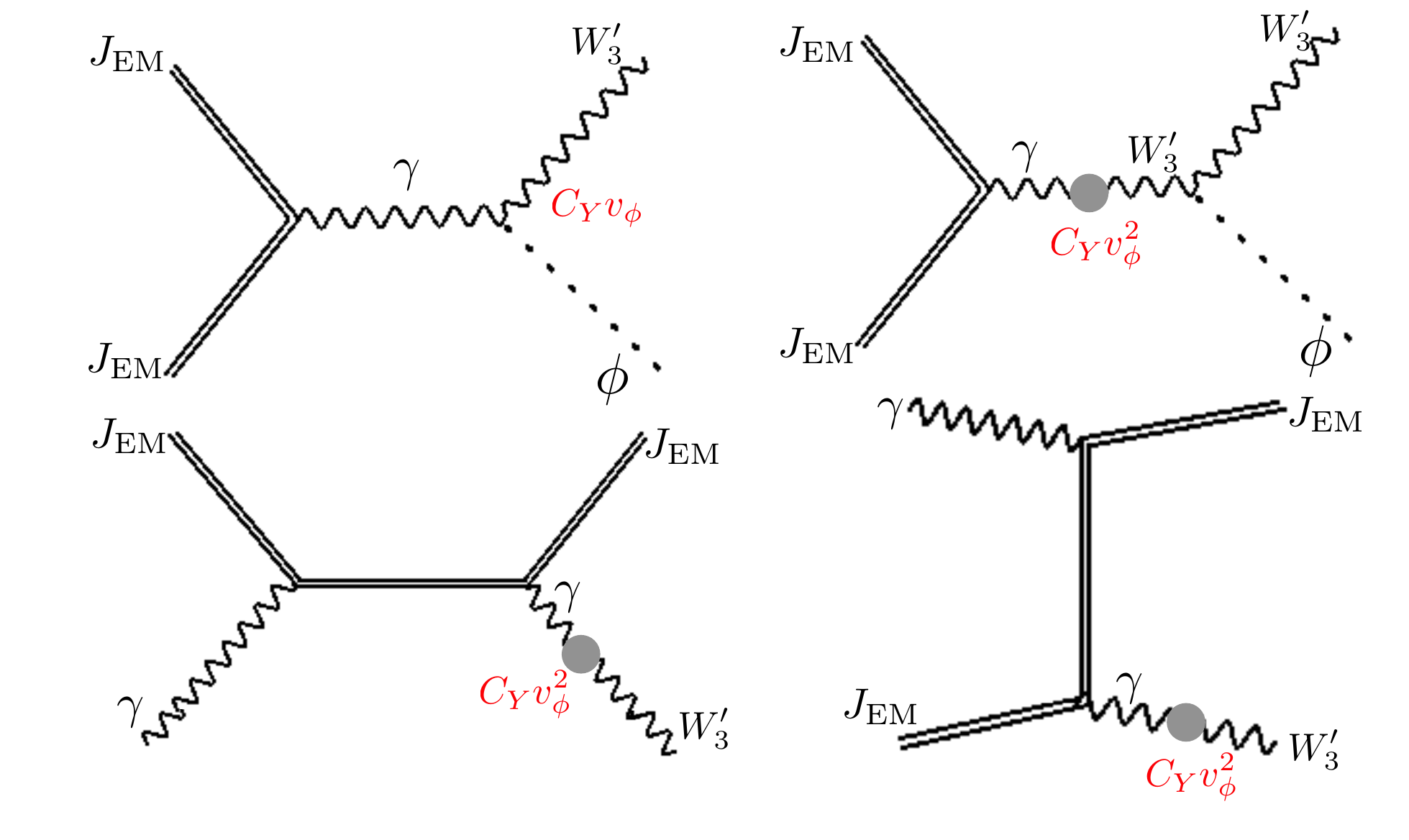}
\caption{ The Feynman diagrams contributing to the production of $W'_3$ and $\phi$ after $\phi$ acquires a \textit{vev}. These diagrams are effective for $T < v_\phi$. In these diagrams, $J_{\text{EM}}$ is any known SM particle that is still around at $T < v_\phi$.  Note that since $v_\phi \simeq 60 \MeV$, the only electromagnetically charged particles that are still in the plasma and have not decayed are the electrons and protons.   }
\label{fig:diagramsW}
\end{figure}

\begin{figure}[h!]
\centering 
\includegraphics[width=0.4\textwidth, height=0.15\textheight]{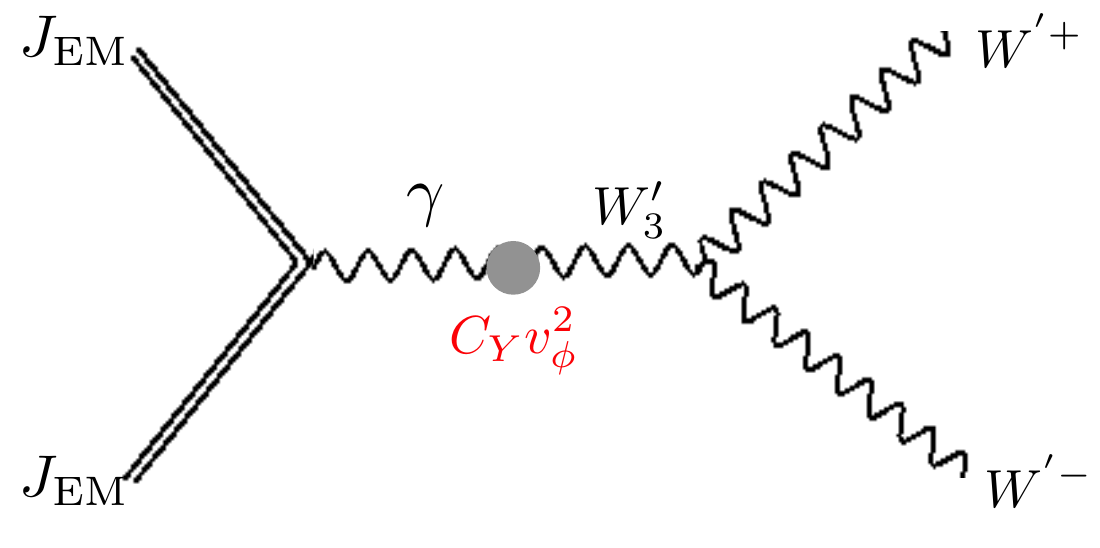}
\caption{The Feynman diagram that shows to the production of $W^{'\pm}$ after $\phi$ acquires a \textit{vev}.}
\label{fig:wpm}
\end{figure}

Once $\phi$ acquires a \textit{vev}, the production of $W'_3$ can occur through renormalizable \textit{and} non-renormalizable operators, shown in Fig.~\ref{fig:diagramsW}. The renormalizable interaction that results in the production of $W^{'\pm}$ is illustrated in Fig.~\ref{fig:wpm}. For processes with two body final states, Eq.~\ref{eq:Bol} simplifies to ~\cite{Elahi:2014fsa}: 
\beq 
\dot n_i + 3 Hn_i =  g_i\frac{3T} {(4 \pi)^4}\int d \Omega  \int_{m_{\text{final}}^2}^{v_\phi^2} ds \left(\frac{s- m_{\text{final}}^2}{\sqrt{s}}\right) |\Mc|^2 K_1\left(\frac{\sqrt{s}}{T}\right),
\label{eq:2to2}
\eeq
where\footnote{The small differences between Eq.~\ref{eq:2to3} and Eq.~\ref{eq:2to2} are due to the number of final state particles, and the fact that final state particles are massive after $SU(2)_D$ SSB. } $m_{\text{final}}$ is the mass of the final state particles in each interaction. The exact value of the squared matrix elements as well as the approximate yield of $W'_3$, $W^{'\pm}$, and $\phi$ can be found in Appendix~\ref{app:ME}. Depending on the benchmark, we can either have both $W^{'\pm}$ and $\phi$, or only $W^{'\pm}$ as our DM particles. However, since the production of $\phi$ and $W^{'\pm}$ are through similar diagrams, these two cases only differ by an $O(1)$ factor.  The region that produces too much DM (i.e, $\Omega_{_{W'}}  + \Omega_{\phi} > \Omega_{_{\text{total DM}}}$) is shown in Orange in Fig.~\ref{fig:bounds}. Even though $\phi$ is long-lived in some regions of the parameter space and can contribute to the relic abundance of DM, this fact does not affect the computations in a significant way. That is because the main production of $\phi$ and $W'$ is due to Eq.~\ref{eq:yieldUV}. Therefore, the bounds are not very sensitive to $m_\phi$. 
  
\subsubsection{CMB and BBN constraints}
 \label{sec:cmb} 

We know that $W'_3$ decays and if it injects energy during BBN, its energetic decay products might disturb the production of the light nuclei by diluting the ration of baryons to photons. Furthermore, the injection of energy may cause photodissociation which will affect the Cosmic Microwave Background (CMB) fluctuations. To avoid these effects, we follow the convention of Ref.~\cite{Cirelli:2016rnw} and require $W'_3$ to decay before it exceeds half of the energy density of the universe. The temperature at which this occurs is 
\beq
T_{\text{dom}} \approx \frac{4 m_{W'}Y_{W'}}{3 f},
\eeq
where $f = 1/2$ is the fraction of the energy density of the universe made up by $W'$. We require that the lifetime of the $W'_3$ is smaller than $H^{-1}(T_{\text{dom}})$. The lifetime of $W'_3$ if $m_{_{W'}} > 2m_e$ is 
\beq
\tau_{_{\text{heavy}}} \simeq \frac{8\pi}{ \alpha C_Y^2 v_\phi m_{_{W'}}} \left(1- \left( \frac{ 2 m_e}{m_{_{W'}}}\right)^2\right)^{-1/2}, 
\eeq
and if it is lighter than $2 m_e$, is 
\beq
\tau_{_{\text{light}}} \simeq  \frac{ 2^7 3^6 5^3 \pi ^3}{17  C_Y^2 v_\phi^4  \alpha^4 m_e} \left(\frac{m_e}{m_{_{W'}}}\right )^9. 
\eeq
  As can be seen in Fig.~\ref{fig:bounds} (shaded blue region), for $m_{_{W'}} < 2 m_e$, this constraint is strongly restricting. However, for $m_{_{W'}} > 2 m_e$, the BBN constraint becomes milder than the bound we get for the relic abundance.  

 
 \subsection{Indirect Detection }
 \label{sec:ID} 
DM accumulating at the Galactic Center or near dwarf spheroidal galaxies, annihilates to $W'_3$: (e.g, $ W^{'+} W^{'-} \to W^{'(*)}_3 W'_3$ ). Depending on the mass of $W'_3$, we may either have $W'_3 \to e^+ e^-$ or $W'_3 \to 3 \gamma$.  An excess emission of positron may be detected by Voyager~\cite{Stone150} and the AMS-02~\cite{PhysRevLett.113.021301}. As discussed in~\cite{Cline:2018ami} and~\cite{1612.07698}, any claim on the detection of DM from the excess positron suffers from large uncertainties and it is not reliable. 

The \textit{Fermi}-LAT collaboration~\cite{1503.02641,Bergstrom:2017ptx} is searching for the excess in photons, and Ref.~\cite{Cline:2018ami} has derived the constraint on DM coming from 6 years of  \textit{Fermi}-LAT observations of 15 dwarf spheroidal galaxies. Ref.~\cite{Cline:2018ami} has shown that the region parameter space satisfying 
$ \langle \sigma v \rangle_{_{W^{'+} W^{'-} \to 3 \gamma  W'_3 \to 6 \gamma}}  > 2.8  \times 10^{-28} \text{cm}^3/\text{s},$
is excluded.  We can approximate this annihilation as
\beq
\langle \sigma v \rangle_{_{W^{'+} W^{'-} \to 3 \gamma  W'_3 \to 6 \gamma}}  \simeq \frac{ \alpha_D^2 C_Y^2 v_\phi^4}{2^8 3^6 5^3 \pi^3 m_{_{W'}}^2  }, 
\eeq
where $\alpha_D = g_D^{2}/(4\pi)$. As can be seen, this cross section is extremely small and does not provide any noteworthy bound on the parameter space. 

 \subsection{Direct Detection and Collider Constraints}
 \label{sec:DD} 
The cross section for $W^{'\pm}$ to scatter on proton with $W'_3$ being the mediator is 
\beq 
\sigma_{e_{W'}} = \frac{8\pi \alpha \alpha_D C_Y^2 v_\phi^4 \mu^2_{e _{W'}}}{(m_{_{W'}}^2+ q^2)^2 },
\eeq
where $\mu_{e _{W'}} \equiv \frac{ m_{_{W'}} m_e}{m_{_{W'}}+ m_e}$ is the reduced mass of the DM-electron system, and $q $ is the momentum transfer between the DM and electron. At electron ionization experiments like SENSEI~\cite{Crisler:2018gci} and XENON10\cite{Bergstrom:2017ptx}, the targeted electrons are usually bound to atoms with typical velocity of a bound electron being $v_e \sim \alpha$. The minimum energy transferred required in these experiments to knock out the bound electron and detect the DM-electron scattering is $q \sim \alpha m_e$. For all of our benchmarks, we have $m_{_{W'}} \gg \alpha m_e$ and thus the momentum transfer can be neglected. For such heavy $W'$, the bounds are rather very mild and they do no provide any noticeable bound on our parameter space. 

\subsubsection{BaBar and \textit{SLAC}}
 \label{sec:collider} 
Another constraint on $C_Y$ comes from the direct production of $W'_3$ and $\gamma$ at E137~\cite{Andreas:2012mt} and {\textit{BaBar}}~\cite{1702.03327} experiments. Ref.~\cite{Cline:2018ami} has worked out this constraints and has found that $C_Y^2 v_\phi^4 < 2.5 \times 10^{-8}$, which means that $C_Y < 6 \times 10^{-4}\GeV^{-2}$ which is much weaker bound than the ones we have discussed so far. Since the coupling of $W'_3$ with $W^{'\pm}$ does not play any role, this constraint is oblivious to the value of $g_D$. The\textit{BaBar} bound is shown as shaded purple in Fig.~\ref{fig:bounds}.

Yet, another important constraint comes from the electron beam dump experiment at \textit{SLAC}~\cite{PhysRevLett.81.1175}, which consists of a $20 \GeV$ electron beam hitting upon a set of fixed aluminum plates. Through the $W'_3-\gamma$, we can create a pair of DM candidates ($W^{'\pm}$): $e N \to e N W^{'*}_3 \to e N W^{'+} W^{'-}$. The DM would then travel through a 179 m hill, followed by 204 m of air and then would be detected by an electromagnetic calorimeter. This process, however, is suppressed. That is because the production of on-shell $W'_3$ is favored, which then would decay back to either $e^+ e^-$ or $3 \gamma$. The large electron-positron pair background coming from the SM photon overwhelms the signal. The \textit{SLAC} experiment requires $C_Y < 0.5 \GeV^{-2}$. 

\subsubsection{Electric Dipole Moment of neutrons}
 \label{sec:EDM} 
 
\begin{figure}[h!]
\centering 
\includegraphics[width=0.4\textwidth, height=0.19\textheight]{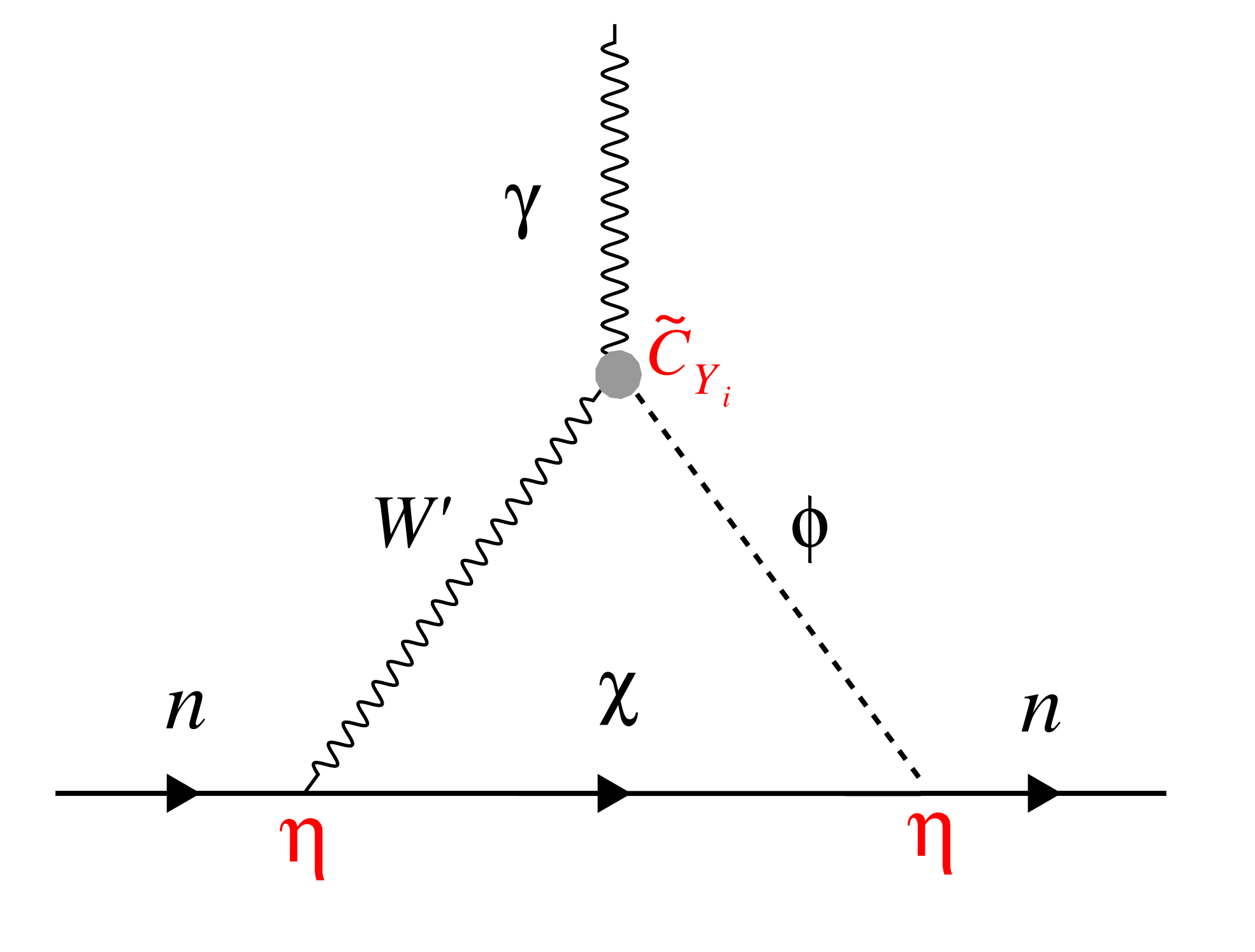}
\caption{ The new contribution to the Electric Dipole Moment of neutron. Since this diagram is suppressed by $\eta^2$, its contribution is very small.}
\label{fig:EDM}
\end{figure}

 Since we can introduce a CP-odd kinetic mixing between the non-abelian fields strength and photon (e.g., $ \tilde C_Y \phi^\dagger\tau^a \phi W^{'\mu\nu}_a F_{\mu \nu}$), we get a constraint on $\tilde C_Y$ from the contribution of this scenario on neutron EDM. The leading contribution is shown in Fig.~\ref{fig:EDM}:\footnote{if we had not turned off $\lambda_{\phi H}$ coupling, we could have an arguably more important contribution to neutron EDM.}
Doing the calculation, we get 
\beq
d_n^{\text{new}} \simeq \frac{\tilde C_{Yi} v_\phi \eta^2}{8\pi^2} \text{log}\frac{\Lambda^2}{m_{W'}^2} 
\eeq
where $\Lambda^2 \sim m_{\Phi_1} m_{\Phi_2}$. Measurements exclude any contribution to the electric dipole moment that exceeds $d_n^{\text{new}} < 10 ^{-26} e. cm$. Given the value of $m_{_{W'}}$ in our benchmarks, EDM measurements require us to $\tilde C_Y < 10^{12} \GeV^{-2}$, and this constraint is much weaker than perturbativity. 
 \subsection{Astrophysical bounds}
 \label{sec:astro} 

\begin{figure}[h!]
\centering 
\includegraphics[width=1\textwidth, height=0.25\textheight]{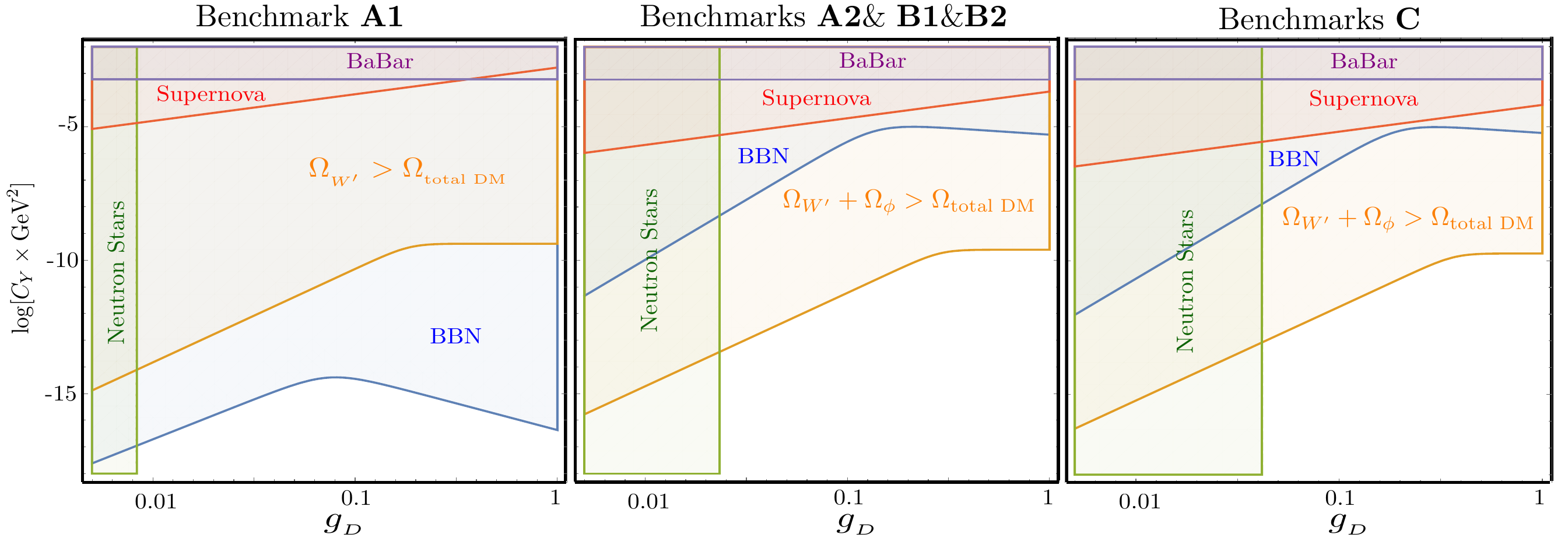}
\caption{ These plots show the constraints from various experiments on different benchmarks. As we can see the most important constraints comes from Relic abundance (Orange), and Neutron Stars (Green)~\cite{Cline:2018ami,1802.08244}. For the case where $m_{_{W'}} < 2 m_e$, BBN~\cite{Cirelli:2016rnw} puts significant restrictions on the parameter space. The bounds coming from \textit{SN1987A}~\cite{Chang:2018rso} and\textit{BaBar}~\cite{1702.03327} are also shown in Red and Purple, respectively. The rest of the bounds discussed are much weaker than these bounds and are not presented here. 
The constraints are very sensitive to the value of $m_{_{W'}}$, but the value of $m_\phi$ does not change the bounds in a visible way. Thereby, we presented the bounds for benchmarks $\Ab, \Ba,$ and $\Bb$ together.  }
\label{fig:bounds}
\end{figure}

We have already summarized the importance of NS in constraining any model that discusses non-standard neutron decay. Recall that to evade NS bounds we moved to models with dark vector mediators and we had to fix $m_{_{W'}}/ g_D \sim 60 \MeV$. This constraint is presented in Fig.~\ref{fig:bounds} as shaded green.

 Another astrophysical bound comes from the cooling rate of Supernova1987A (\textit{SN1987A})~\cite{Chang:2018rso}. Through the mixing with photon, DM can be produced through the implosion of a newly born NS. Since DM does not interact with baryonic matter strongly, it can leaves the supernova, resulting in a faster cooling rate. If DM is produced in appreciable number, then the cooling rate can be faster than observed. For \textit{SN1987A}, the energy loss per unit mass should be smaller than $10^{19}\rm erg/g/s$ at the temperature of the plasma, which equates roughly $10 \MeV$. The shaded red region in Fig.~\ref{fig:bounds} illustrates the constraint coming from \textit{SN1987A}. 
 
 \section{Conclusion}
 \label{sec:conclusion} 

 In this paper, we presented a model that can explain the discrepancy between the total decay width of the neutron and its decay width to protons. In the Standard Model (SM), we expect the branching ratio of $n \to p e^- \bar v_e$ to be $100 \%$. However, the two bottle experiment and beam experiment which tried calculating the decay width of the neutron, one by counting the remaining neutron and another by counting the produced protons, show  a discrepancy in their results. One potential answer could be that neutrons decay to dark sector (DS) particles with a branching ratio of $1\%$. The observation of large neutron stars with radius of two solar radius leads us to only consider DS models with vector mediators to ensure a repulsive interaction between dark matter (DM) candidates as well as between DM and neutrons. A dark $U(1)_D$ gauge has already been discussed in details and has been shown the resulting free parameter space is very small. The important constraints on this scenario comes from the measurements of Cosmic Microwave Background (CMB) and Big Bang Nucleosynthesis (BBN) which strongly disfavor the existence of a light degree of freedom in large abundance at late times. To avoid these constraints, we considered the production of DS through freeze-in mechanism. Even with freeze-in, however, we showed that the region of the parameter space that explains the neutron decay anomaly will necessarily lead to the over production of $\chi$-- the fermionic DM candidate in our theory. Thereby, we considered a low  $\Tmax$( e.g, $T_{\text{max}} \sim m_\chi$). This temperature is valid according to the current constraints on the reheat temperature of the universe. 
 
Since $\chi$ in this model cannot account for the relic abundance of DM in the early universe, we considered a DS with gauged $SU(2)_D$. The extra degrees of freedom in this model can successfully account for the observed relic abundance of DM. Yet, the number of free parameters in this model is very much like gauged $U(1)_D$, due to the intricate relationship between the particles of DS.  
 
 One important advantage of DS scenarios that attempt to explain another theoretical or experimental anomaly is that the freedom over the new parameter space becomes much smaller. In this paper, we only had a few free parameters we could play with: the kinetic mixing coupling, the dark gauge coupling and $m_{_{W'}}$ which could vary over a small region. For $m_{_{W'}} > 2 m_e$, satisfying the right relic abundance gave the best bound on the kinetic mixing between the two sectors.  for lighter $W'_3$ ($m_{_{W'}} < 2 m_e$), BBN constraints became much more significant. The main constraint on $g_D$ is from making sure the self interaction of DM, as well as the interaction between DM and neutrons are repulsive enough that they do not change the equation of state of large neutron stars significantly. 
 
\acknowledgments 
We would like to thank H. Mehrabpour and J. Unwin for numerous useful conversations. FE is also thankful to CERN theory division and Mainz Cluster of Excellence for their hospitality. 

 \appendix
 \section{The squared Matrix Elements of the processes that produce $W'$ and $\phi$ for $T < v_\phi$.}
 \label{app:ME} 
 
The exact Matrix Element of the processes presented in Fig.~\ref{fig:diagramsW} and Fig.~\ref{fig:wpm} are the following:
\begin{align*}
|\Mc|^2_{_{ J_{_{\text{EM}}} J_{_{\text{EM}}} \to \phi W'}}\hspace{0.25 in} &= \frac{Q^2 C_Y^2 v_\phi^2 }{4}(g_D v_\phi  + \sqrt{s} )^2 \frac{2 m_{_{W'}}^2 ( 4 s^2 + 5s (t+u) - 5tu) + 5s (s-t)^2}{m_{_{W'}}^2 s^2} \\
|\Mc|^2_{\gamma_{ J_{_{\text{EM}}} \to J_{_{\text{EM}}} W'}} \hspace{0.25 in}&= \frac{Q^4 C_Y^2 v_\phi^4}{4} \frac{ ( 2 m_{_{W'}}^2  s(2 s-t+u)+s^2 (2 s-t-5 u))}{m_{_{W'}}^2 (s-t)^2} \\
|\Mc|^2_{_{J_{_{\text{EM}}} J_{_{\text{EM}}} \to W^{'+} W^{'-}}} &= \frac{g_D^2 Q^2 C_Y^2 v_\phi^4}{32 m_{_{W'}}^4 \left(s-m_{_{W'}}^2\right)^2}\left(32 m_{_{\text{SM}}}^4 (m_{_{W'}}^2 s-4 m_{_{W'}}^4)\right.\\
&+2 m_{_{\text{SM}}}^2 (s-4 m_{_{W'}}^2) (s^2+4 m_{_{W'}}^2 (s+4 (t+u))-20 m_{_{W'}}^4)-128 m_{_{W'}}^8\\
&-16 m_{_{W'}}^6 (s-8 (t+u))-4 m_{_{W'}}^4 (s^2+8 s (t+u)+11 t^2+10 t u+11 u^2)\\
&\left.+4 m_{_{W'}}^2 s (3 t^2+2 t u+3 u^2)+s^2 (s^2- (t-u)^2)\right),  \\
\end{align*}
where the Mandeslestam ($s,t, u$) variables are defined as usual: 
\begin{align*}
s &=  4 (T^2 +  m_{_{\text{SM}}}^2)\\
t&= - 2(T^2 +  m_{_{\text{SM}}}^2)  +2 T \sqrt{ T^2 + m_{_{\text{SM}}}^2 - m_{_{W'}}^2} ( \cos \theta) +  m_{_{\text{SM}}}^2 + m_{_{W'}}^2\\
u&=- 2(T^2 +  m_{_{\text{SM}}}^2)  -2 T \sqrt{ T^2 + m_{_{\text{SM}}}^2 - m_{_{W'}}^2} ( \cos \theta) +  m_{_{\text{SM}}}^2 + m_{_{W'}}^2.\\
 \end{align*}
 
In the limit where $s \gg t, u$ and $ v_\phi >  \sqrt{s} >  m_{_{\text{SM}}}^2 (T=0)$, we get 
\begin{align*}
|\Mc|^2_{_{ J_{_{\text{EM}}} J_{_{\text{EM}}} \to \phi W'}}\hspace{0.25 in} &\simeq \frac{ Q^2 C_Y^2 v_\phi^2 (g_D v_\phi  + \sqrt{s} )^2  s }{ m_{W'}^2}  \\
|\Mc|^2_{\gamma_{ J_{_{\text{EM}}} \to J_{_{\text{EM}}} W'}} \hspace{0.25 in}&\simeq 2\frac{ Q^4 C_Y^2 v_\phi^4 \ s}{m_{_{W'}}^2} \\
|\Mc|^2_{_{J_{_{\text{EM}}} J_{_{\text{EM}}} \to W^{'+} W^{'-}}} &\simeq \frac{g_D^2 Q^2 C_Y^2 v_\phi^4 s^2}{8 m_{_{W'}}^4 }.
\end{align*}

Given that for all of our benchmarks $v_\phi$ is at least an order of magnitude greater than $m_{_{W'}}$ and $m_\phi$, we can ignore $m_{\text{final}}$ in some of the cases of our interest. The yield, thus, becomes the following:
\begin{align*} 
Y^{^{T< v_\phi}}& = Y_{_{ J_{_{\text{EM}}} J_{_{\text{EM}}} \to \phi W'}}+ Y_{_{ \gamma J_{_{\text{EM}}}\to  J_{_{\text{EM}}} W'}}+ Y_{_{ J_{_{\text{EM}}} J_{_{\text{EM}}} \to W' W'}},\\
Y_{_{ J_{_{\text{EM}}} J_{_{\text{EM}}} \to \phi W'}}& \simeq  g_i \alpha M_{Pl} \frac{45 \times C_Y^2  }{ 1.66 \times 16  \pi \sqrt{g_\star^{\rho}} g_\star^S} \frac{v_\phi^5}{m_{_{W'}}^2} \left[ 64 g_D \left(1- \frac{m_{\text{final}}}{v_\phi}\right)   +  45 \pi \left(1- \left( \frac{m_{\text{final}}}{v_\phi}\right)^2\right)\right]^2,\\
Y_{_{ \gamma J_{_{\text{EM}}}\to  J_{_{\text{EM}}} W'}}& \simeq g_i \alpha M_{Pl} \frac{ 45 \times C_Y^2 }{1.66\times 4  \pi \sqrt{g_\star^{\rho}} g_\star^S}\frac{v_\phi^5}{m_{_{W'}}^2} \left[ 1 - \frac{m_{\text{final}}}{v_\phi} \right]\\
Y_{_{ J_{_{\text{EM}}} J_{_{\text{EM}}} \to W' W'}} & \simeq \alpha  M_{Pl} \frac{45 * 2^5 \times C_Y^2 }{ 1.66 \times  \pi \sqrt{g_\star^{\rho}} g_\star^S} \frac{v_\phi^7}{m_{_{W'}}^4}  \left[ 1 - \left(\frac{m_{\text{final}}}{v_\phi}\right)^3 \right],
\end{align*}

 \bibliography{neutron_decay}
\end{document}